\documentclass[acus]{JAC2000}

\usepackage{graphicx}

\begin{document}
\title{\flushright{THAT004}\\[15pt] \centering DISTRIBUTED CONTROL SYSTEM
  FOR THE TEST INTERFEROMETER OF THE ALMA PROJECT}

\author{M.~Pokorny, M.~Brooks, NRAO, Tucson, AZ, USA\\
  B.~Glendenning, G.~Harris, R.~Heald, F.~Stauffer, NRAO, Socorro, NM,
  USA\\
  J.~Pisano, NRAO, Charlottesville, VA, USA}

\maketitle

\begin{abstract}
  The control system (TICS) for the test interferometer being built to
  support the development of the Atacama Large Millimeter Array
  (ALMA)\cite{alma-web} will itself be a prototype for the final ALMA
  array, providing a test for the distributed control system under
  development. TICS will be based on the ALMA Common Software
  (ACS)\cite{acs-paper} (developed at the European Southern
  Observatory), which provides CORBA-based\cite{corba-web} services
  and a device management framework for the control software.
  
  Simple device controllers will run on single board computers, one of
  which (known as an LCU) is located at each antenna; complex,
  compound device controllers may run on centrally located computers.
  In either circumstance, client programs may obtain direct CORBA
  references to the devices and their properties.  Monitor and control
  requests are sent to devices or properties, which then process and
  forward the commands to the appropriate hardware devices as
  required.  Timing requirements are met by tagging commands with
  (future) timestamps synchronized to a timing pulse, which is
  regulated by a central reference generator, and is distributed to
  all hardware devices in the array. Monitoring is provided through a
  publish/subscribe CORBA-based service.
\end{abstract}

\section{ALMA}

The Atacama Large Millimeter Array, or ALMA, is a radio astronomy
millimeter and sub-millimeter array to be built in Chile's Atacama
desert (at an elevation of 5000m above sea level) in the coming years.
The project is an international collaboration among partners from
Europe, Japan, Canada, Chile, and the USA. Current plans are for an
array consisting of sixty-four parabolic antennas of twelve meter
diameter, with configurations ranging in size from a compact
configuration, with a maximum baseline of 150 m, to an extended
configuration, with a maximum baseline of 10 km. Each of the antennas
will contain from four to ten cryogenically cooled receivers,
operating in the range from 31 GHz to 950 GHz.  Science data will be
digitized at each antenna, and transmitted via optical fiber to a
central location at a rate of 3 GB/s per antenna.

\section{Test Interferometer Control System: TICS}

Three antennas, one contracted by each of the high-level partners in
the ALMA project, will be delivered over the next two years to the
National Radio Astronomy Observatory's Very Large Array site in New
Mexico. These antennas are collectively known as the Test
Interferometer (TI). The TI will be used for comparative evaluations
of the three products to assess which design will be used for the
antennas in the final ALMA array.  Simultaneously, the TI will also be
employed as a test-bed for the development of technologies to be used
by the ALMA array; in particular, the development of the control
system software for the ALMA project. Nonetheless, the primary goal of
TICS is to provide the TI itself with a comprehensive array control
system.

\subsection{ALMA Common Software: ACS}

The common base of all software being developed for ALMA will be the
ALMA Common Software (ACS). Most of the development of ACS is being
done by the European Southern Observatory, based in Garching, Germany.
ACS provides a CORBA-based device management framework, system
services, application services, and an application framework for all
software that will be required for ALMA. TICS is the first
application to use ACS, and has been developed roughly in parallel
with ACS. The device management framework and system services have
been the primary features used in the development of TICS to this
point in time. Some of the ACS features employed by TICS include:
\begin{Itemize}
\item object lifetime management
\item configuration database (devices and properties)
\item naming service
\item time service.
\end{Itemize}

\subsection{Control System Architecture}

At each antenna there is an Antenna Bus Master (ABM): a VME bus Power
PC based computer running the VxWorks operating system. Its principal
role is to provide real-time control of the devices at the antenna
based upon infrequent time-tagged commands from the center. The ABM
also serves as a router for an antenna Ethernet segment.

Most devices with computer interfaces are attached to a Controller Area
Network (CAN) bus, through which they are controlled and
monitored by the ABM.

Each ABM is connected to the central systems via a point-to-point
Gigabit Ethernet network that terminates at a switch. The switch is in
turn connected to a high-speed switched network on which all central
ALMA computer systems required to operate the array are attached.

Two real-time computers are situated at a central location of the
array. The Array Real-Time Machine (ARTM) plays the role of the ABM at
the central location, providing local real-time control of its
attached devices. The other central real-time computer, the
Correlator Control Computer (CCC), provides the interface for the
correlator, and detailed control of the correlator hardware. Both the
ARTM and CCC are VME/PPC/VxWorks based systems.

The coordination function is implemented via the Array Control
Computer (ACC), which is a high-end workstation running the Linux
operating system. It is responsible for controlling all hardware in
the array (indirectly through the ABM, ARTM, and CCC computers) under
the command of a high-level observing script. The ACC also runs
various ancillary software such as model servers (\textit{e.g.}, phase
models), and data formatting.

Almost all devices will be attached to a CAN bus operating in a
master/slave (polled) fashion. The bus will operate at 1 Mbps and is
capable of at least 2000 polled operations per second (up to 8 bytes
of data per transaction). Devices on the CAN bus will be responsible
for implementing a simple in-house protocol to map CAN message IDs to
internal device addresses. A few devices will have other connections,
in particular Ethernet.

\subsection{Devices}

Logically, the software is partitioned so that control flows in a
master-slave fashion from a central executive, which controls
high-level (``composite'') software devices, which in turn control
their constituent parts. The lowest level software devices are
referred to as device controllers, and represent a proxy for the
actual hardware --- that is, they communicate with the hardware. Data,
both monitor and back-end, are collected from the devices by a
collecting process in the real-time computer connected to the
hardware, and are then buffered up for distribution, {\it via} a
publish/subscribe mechanism, to data consumers. Standard data
consumers include processes that format and archive the data. The
software is distributed among the computers so that only the device
controllers and software directly concerned with low-level device
activities are on the local real-time computer.  All higher-level
software entities are concentrated on the ACC.

Engineering access to devices that are installed on the test
interferometer will be implemented by access to the device controller
interface, or to the I/O routines directly from engineering
workstations.

Naming services provided by ACS are used by clients to obtain CORBA
references to the devices. Transient servants are created and
destroyed by the ACS ``Manager'' as they are needed by clients, but
persistent servants (typically, those devices closest to the hardware)
can also be created at system start-up. Device configuration and some
aspects of system configuration are implemented using a centralized
database.

\subsection{Properties}

Devices in the control system have both properties and methods. Since
properties, like devices, are themselves CORBA objects, clients may
get references to properties or devices that are physically located on
any computer on the network. Using features of ACS, properties may
have monitors or alarms attached to them, or a client may simply poll
a property at will.

\subsection{Timing}

The ALMA time system will establish synchronized switching cycles and
mode changes, and provide time-stamping for the resulting measurements
across the entire instrument, including the central building and the
geographically dispersed antennas. Additionally, the time system must
be accurately related to external measures of time to correctly
determine the position of astronomical objects of interest. The
fundamental time system of the interferometer is TAI time maintained
in a central master clock.

While most devices do not have precise timing requirements, a timing
pulse with a 48 ms period is distributed throughout the entire array
to provide a time basis for those devices that do have more precise
timing requirements. For such a device, the control software must
arrange to have monitor and control commands sent to the device in
precisely defined windows within the 48 ms timing period. Time-tagged
commands from the center must be transmitted sufficiently early to
account for the non-determinism in the network and general-purpose
ACC. The slave clocks are given the array time of a particular timing
event, and thereafter maintain time by counting timing events.

Some of the relevant time scales in the control system are as follows.
\begin{Itemize}
\item 2 ms: The shortest time-scale at which any device will require
  interaction. Shorter time-scales are always handled by hardware.
\item 16 ms: Fastest correlator dump time.
\item 48 ms: The period of the pervasive timing event sent to
  all hardware with precise timing requirements.
\item 1 s: The fastest time-scale for observational changes,
  \textit{e.g.}, source changes or changes in correlator setup.
\item $>$1 s: Most devices will be monitored or controlled at rates
  slower than 1 Hz, often much slower (300 s).
\end{Itemize}

\section{Example Device: Fine Tuning Synthesizer}

The fine tuning synthesizer (FTS) is a low-level device that is a
component of local oscillators throughout the array. It provides the
fine adjustment of the local oscillator phase and frequency (for the
purposes of fringe tracking), and phase switching capabilities (used
to remove spurious signals and for sideband separation). Like most
devices, the FTS hardware is monitored or controlled \textit{via} the
antenna-wide CAN bus.

To present an example of the execution of TICS, a short description of
some of the properties of the FTS is given here.

\subsection{Timing}

The FTS implements several timing event associated commands. All of
the functionality of the FTS depends upon synchronization with the
pervasive timing event. Effective phase switching requires
synchronization between antennas at the start of the phase switching
function. Fringe tracking depends upon tracking and pointing
information to achieve the desired phase output at the proper time;
therefore, updates to the phase function must remain synchronized with
the timing event.

The time scales of importance to the FTS are the following:
\begin{Itemize}
\item 250 $\mu$s: shortest phase switching interval
\item 16 ms: fast phase switching period, shortest slow phase switching
  interval
\item 48 ms: phase chirp modulation update rate
\item $\sim$1 s: slow phase switching period
\item $\sim$10 s: fast switching calibration
\item $\sim$100 s: fringe tracking frequency update.
\end{Itemize}

\subsection{Properties}

\subsubsection{Fringe tracking}

Local oscillators will be compound devices, each containing an FTS
device. A command from a high-level device to a local oscillator
device to set a frequency will occur at the array network level
through CORBA. Such a command will occur whenever the astronomical
source or receiving frequency is changed.

The inputs required by the FTS for fringe tracking are all dynamic,
depending upon factors such as the observing frequency and tracking
information. A client of the FTS device, namely a local oscillator,
will be responsible for setting these properties as necessary. On the
other hand, the phase function needed by the FTS to provide accurate
phase tracking services may be obtained by the local oscillator from a
phase model server running on the ACC.

The FTS device controller will send commands to the hardware on the
CAN bus just prior to the timing event on which the changes are to
take effect. After setting initial conditions, the FTS device may then
update a phase chirp modulation parameter at the 20 5/6 Hz rate to
adjust for non-linearities in the phase function, and to control
quantization errors in the phase generation.

\subsubsection{Phase switching}

For phase switching, each FTS uses a four-valued step function with a
minimum interval of 250~$\mu$s between changes, and a period of
1.024~s. The functions comprise a mutually orthogonal set, from which
each antenna uses a single element over its ``lifetime''. There are in
fact two, nested two-state switching cycles, which compose the overall
1.024 s cycle; however, the functions used by the two cycles are
similar, differing only in scaling on the time axis, and a shift on
the phase axis.  The phase switching function to be used by a
particular FTS may therefore be configured using a value from the
configuration database, allowing the device controller to set its own
phase switching function when it is instantiated.

Because the rate of phase switching is faster than it is possible to
accurately control from the ABM, the phase switching itself is
implemented by the FTS hardware. The precision required of the phase
switch times is ensured by an especially precise ``version'' of the 48
ms timing pulse that is received by the FTS hardware.

\section{TICS Status}

The first version of TICS (version 0.1) was released earlier this
year, and the next release is scheduled to occur in February of next
year. By mid-April of 2002, the first TI antenna will be ready for
evaluation. TICS will support mount control and pointing tests by the
time the antenna is delivered. As the antenna proceeds through the
evaluation process, TICS will provide new capabilities to support
required testing. Concurrently, further integration with ACS and use
of new ACS features (as they are developed) will also occur.

\end{document}